\documentclass{moriond}

\def\be{\begin{equation}}
\def\ee{\end{equation}}
\def\bea{\begin{eqnarray}}
\def\eea{\end{eqnarray}}

\begin{document}
\vspace*{4cm}
\title{Managing the Growing Complexity of Multi-Messenger Transient Events with Astro-COLIBRI}

\author{ Fabian Schüssler\\ A. Kaan Alkan, M. de Bony de Lavergne, and J. Mourier}
\address{IRFU, CEA, Université Paris-Saclay, F-91191 Gif-sur-Yvette, France}

\maketitle\abstracts{
Observations of transient phenomena, such as GRBs, FRBs, novae/supernovae explosions, coupled with the detection of cosmic messengers like high-energy neutrinos and gravitational waves, have transformed astrophysics. Maximizing the discovery potential necessitates tools for swiftly acquiring an overview of the most relevant information for each new detection. Introducing Astro-COLIBRI, a comprehensive platform designed to meet this challenge. Astro-COLIBRI features a public API, real-time databases and alert systems, a discussion forum, and a website and iOS/Android apps as user clients. In real time, it evaluates incoming astronomical observation messages from all available alert streams, filters them based on user-defined criteria, and contextualizes them in the multi-wavelength (MWL) and multi-messenger (MM) context. User clients offer a graphical representation, providing a succinct summary for quick identification of interesting phenomena and assessing observing conditions globally.}


\section*{Introduction}
Astro-COLIBRI~\cite{2021ApJS..256....5R} is a state-of-the-art platform designed to enhance the study of transient astronomical phenomena by integrating real-time multi-messenger observational tools within a unified and accessible interface. This platform serves both professional and amateur astronomers, offering a comprehensive framework for the coordination of follow-up observations, thereby maximizing the scientific return from transient events. By aggregating alerts and data from a multitude of observatories and detection systems, Astro-COLIBRI enables the rapid identification and thorough analysis of phenomena such as Gamma-Ray Bursts (GRBs), Fast Radio Bursts (FRBs), supernovae, and gravitational waves.

\begin{figure}[t!]
\begin{center}
\includegraphics[width= 0.75\textwidth]{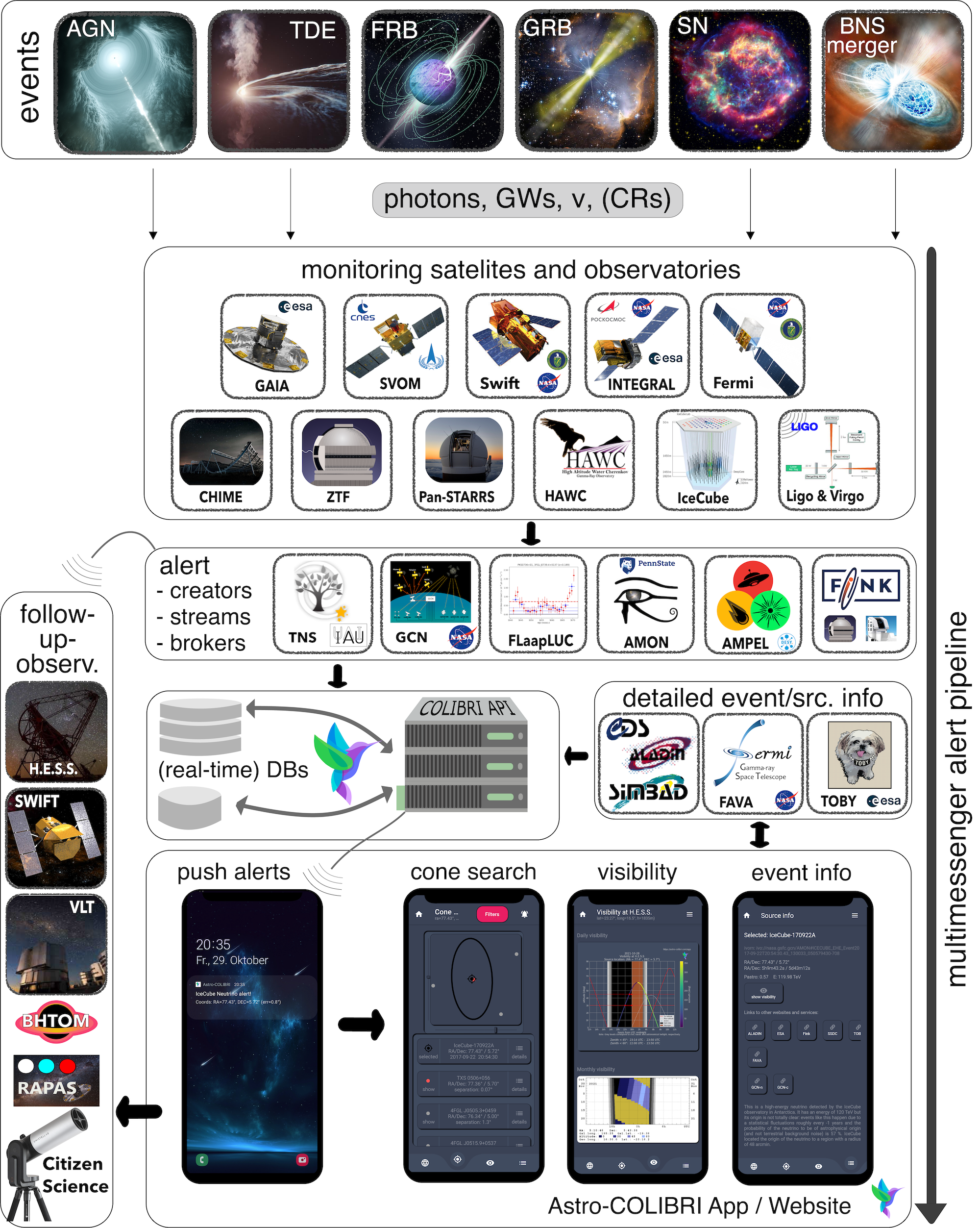}
\caption{Astro-COLIBRI is conveived as top-level platfrom fully integrated in the global multi-wavelength and multi-messenger landscape.}
\label{fig:landscape}
\end{center}
\end{figure}

The fundamental objective of Astro-COLIBRI is to streamline the observation process for transient events, which are often brief and require immediate attention. By providing real-time alerts and detailed contextual information, the platform empowers astronomers to respond promptly and effectively to new discoveries. The integration of multi-wavelength and multi-messenger data enhances the understanding of these events, facilitating more comprehensive studies and fostering collaborative research efforts. The place of Astro-COLIBRI in the global time domain and multi-messenger landscape is depicted in Fig.~\ref{fig:landscape}.

Astro-COLIBRI's architecture features a public RESTful API, real-time databases, a cloud-based alert system, and both web and mobile applications. This infrastructure supports a wide spectrum of astrophysical phenomena, equipping users with the tools necessary to plan and execute observations across various wavelengths and messengers. The platform's intuitive graphical interfaces, available on iOS and Android devices, ensure broad accessibility, engaging both professional and citizen scientists who are integral to the observation network.

\section*{Overview of core features}

Astro-COLIBRI was developed to address the escalating need for a cohesive system capable of managing the intricate requirements of transient event observation. The platform's core features are meticulously designed to provide extensive support for multi-messenger astrophysics:

\begin{enumerate}
    \item \textbf{Real-Time Alerts and Notifications}: Astro-COLIBRI continuously processes incoming messages from diverse alert streams like the Global Coordinates Network (GCN) and the Transient Name Server (TNS) in real-time. A large variety of dedicated notification streams ensures timely notifications of relevant events. This capability facilitates prompt decision-making and subsequent observational activities.
    
    \item \textbf{Graphical User Interfaces}: The platform presents user-friendly web and mobile interfaces that offer a comprehensive overview of transient events, including their spatial locations, classifications, and pertinent observational data. Users can customize display settings and filter options to tailor the interface to their specific requirements.
    
    \item \textbf{Observatory Selection and Scheduling}: Astro-COLIBRI provides advanced tools for selecting observatories and verifying observation conditions. Users can specify custom observer locations, which include both professional and amateur observatories, and generate optimized observation plans for events such as gravitational wave detections.
    
    \item \textbf{Multi-Wavelength and Multi-Messenger Integration}: The platform integrates data from multiple sources, including high-energy neutrinos, optical transients, gamma-ray bursts, and gravitational waves. This integration facilitates detailed event analysis and enhances the potential for groundbreaking discoveries through coordinated multi-messenger observations.
    
    \item \textbf{API Access}: The public RESTful API allows for seamless integration of Astro-COLIBRI's data into external systems and tools. This flexibility supports a broad range of applications, from professional observatories to amateur astronomer setups, thereby broadening participation in transient event studies.
    
    \item \textbf{Citizen Science Support}: Recognizing the significant contributions of amateur astronomers, Astro-COLIBRI includes features specifically designed for their needs. These features encompass notifications for bright events and tools for defining custom observatory locations, thereby fostering greater engagement from the citizen science community and strengthening the overall observational network.
\end{enumerate}


\section*{Event Filtering and Selection Tools in Astro-COLIBRI}

Astro-COLIBRI incorporates sophisticated tools to filter and select transient events, ensuring that users can efficiently identify the most relevant phenomena for their research needs. The platform facilitates a top-level selection process based on the observatory that detected the transient event and the type of event itself, such as Gamma-Ray Bursts (GRBs), supernovae (SN), or gravitational waves (GWs).

These top-level selections are implemented as intuitive, easy-to-use buttons within the Astro-COLIBRI interfaces. Users can quickly filter events by simply (de-)activating these buttons, representing various observatories and event types. This streamlined approach allows for rapid initial filtering, enabling users to focus on the most pertinent events.

For users requiring more specific criteria, Astro-COLIBRI provides additional, more detailed filtering options. By long-pressing on the top-level selection buttons, a submenu is opened, revealing a scrollable list of filter parameters. The values used for the filtering are user defined and are logically combined to offer a comprehensive and customizable filtering experience. 

The detailed filters include a wide range of parameters such as:
\begin{itemize}
    \item event magnitudes at the time of detection of optical transients
    \item signalness and event types for high-energy neutrinos
    \item significance, localisation uncertainty, event classification (e.g. probablity for BBH, BNS, NS-BH, MassGap, etc.) for gravitational waves
\end{itemize}

\begin{figure}[t!]
\begin{center}
\includegraphics[width= 0.75\textwidth]{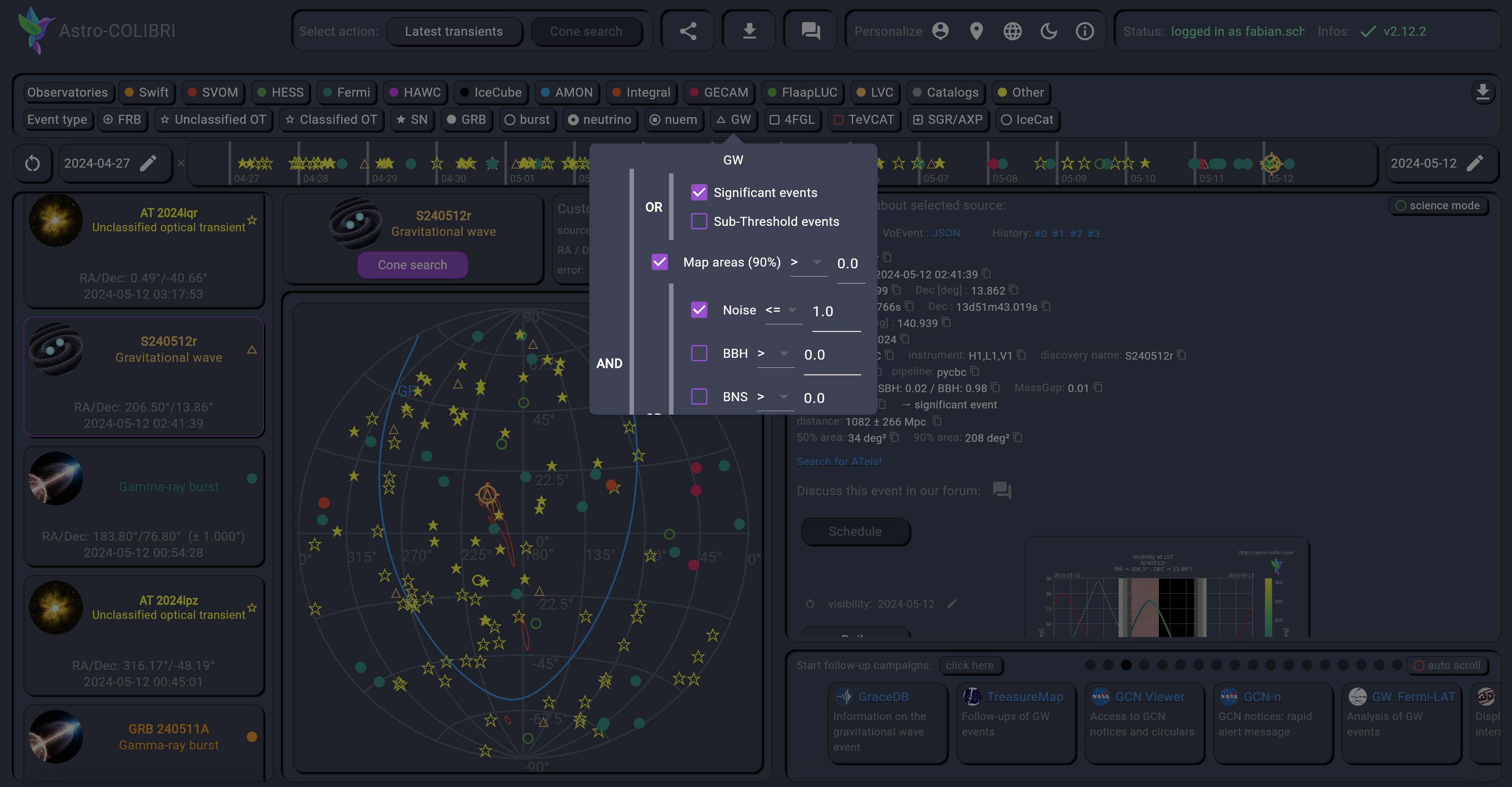}
\caption{Example of detailed event filters available in the Astro-COLIBRI interfaces by long-pressing the relevant top-level filter button.}
\label{fig:gw}
\end{center}
\end{figure}

An example of the filters for GWs is shown in Fig.~\ref{fig:gw}. This granular level of control ensures that users can tailor their event selection to match their specific observational goals and instrumentation capabilities. The logically combined filters enhance the platform's flexibility, making it a powerful tool for both broad surveys and targeted studies.

By providing both top-level as well as detailed filtering options, Astro-COLIBRI empowers astronomers to efficiently navigate the increasing complexity and volume of transient event data. This capability is essential for maximizing the scientific return from transient event observations and for enabling timely and coordinated follow-up studies.

\section*{Conclusion}

Astro-COLIBRI represents a significant advancement in the field of multi-messenger astrophysics, offering a comprehensive and efficient solution for the real-time observation and analysis of transient astronomical events. By consolidating various data sources and providing user-friendly interfaces, the platform enables both professional and amateur astronomers to deepen their understanding of the dynamic universe. The collaborative potential facilitated by Astro-COLIBRI enhances the overall scientific output and promotes a more interconnected and informed astronomical community.

The rising number and complexity of transient events present considerable challenges to the astronomical community. Astro-COLIBRI addresses these challenges by providing a centralized platform capable of managing and interpreting the extensive data generated by these phenomena. The ability to swiftly and accurately identify, classify, and respond to transient events is essential for advancing astrophysical research. As new observatories come online and detection technologies evolve, Astro-COLIBRI's role in managing this data influx will become increasingly critical, ensuring that astronomers can effectively navigate the rapidly changing landscape of transient event research.


\section*{Acknowledgements}
The authors acknowledge the support of the French Agence Nationale de la Recherche (ANR) under reference ANR-22-CE31-0012. This work was also supported by the Programme National des Hautes Energies of CNRS/INSU with INP and IN2P3, co-funded by CEA and CNES and we acknowledge support by the European Union’s Horizon 2020 Programme under the AHEAD2020 project (grant agreement n. 871158).

\end{document}